\documentclass[aps,showpacs,prl,
reprint,
  ]{revtex4-1}
\usepackage{graphicx}
\usepackage{natbib}
\usepackage{amsmath}
\usepackage{epstopdf}
\usepackage{color}
\usepackage{upgreek}
\usepackage{gensymb}
\begin{document}

\title{Twin-beam-enhanced displacement measurement of a membrane in a cavity}

\author{Xinrui Wei$^1$}
\author{Jiteng Sheng$^{1,*}$}
\author{Yuelong Wu$^1$}
\author{Wuming Liu$^2$}
\author{Haibin Wu$^{1,\dag}$}

\affiliation{$^1$State Key Laboratory of Precision Spectroscopy, East China Normal University, Shanghai 200062, China}
\affiliation{$^2$Beijing National Laboratory for Condensed Matter Physics and Institute of Physics, Chinese Academy of Sciences, Beijing 100190, China}


\begin{abstract}
Ultrasensitive measurement of a small displacement is an essential goal in various applications of science and technology, ranging from large-scale laser interferometric gravitational wave detectors to micro-electro-mechanical-systems-based force microscopy. The least measurable displacement is ultimately limited by the quantum nature of light in a classical optical sensor. Here we use the bright quantum correlated light, i.e., twin beams, generated by a coherent atomic medium to surpass the shot-noise limit (SNL) of the displacement measurement of a membrane in an optical cavity. The sensitivity of 200 $am/\sqrt {Hz}$ is achieved, which is more than two orders of magnitude better than a standard Michelson interferometer. An improvement of 3 $dB$ in the signal-to-noise ratio (SNR) beyond the SNL is realized at an equivalent optical power, by using quantum correlated light with noise squeezed 7 $dB$ below the vacuum level. Moreover, the frequency correlation of twin beams is directly measured by using optical cavities, and this relation is utilized to reduce the excess classical noise. Additionally, the displacement measurement sensitivity is further substantially enhanced by the cavity mediated dispersion and the SNR is increased by one order of magnitude compared to the free space case. These results provide a novel strategy for the world of precision measurement as well as to control cavity optomechanical systems with non-classical light.
\end{abstract}


\maketitle

\section{I. Introduction}

It has been long known that non-classical state of light can be utilized to surpass the shot-noise limit (SNL) and improve the measurement sensitivity of an optical sensor \cite{bachor2004guide}. To implement such a measurement beyond the SNL, a typical way is to inject a single-beam quadrature-squeezed state of light into an interferometer and measure the signal combined with a local oscillator via a balanced homodyne detector \cite{PhysRevD.23.1693,PhysRevLett.59.278,PhysRevLett.59.2153,PhysRevLett.104.251102,PhysRevLett.88.231102,PhysRevLett.95.211102,goda2008quantum,aasi2013enhanced}. One of the most impressive examples is the enhanced sensitivity of the Laser Interferometer Gravitational-Wave Observatory (LIGO) detector with squeezed light, resulting in an improvement of 2.15 $dB$ by injecting 10.3 $dB$ of squeezing \cite{aasi2013enhanced}.

In this work, we utilize an alternative quantum light source, i.e., quantum correlated light or twin beams, and demonstrate that the sub-shot-noise and highly sensitive measurement of a silicon nitride (SiN) membrane's displacement can be realized with quantum correlated light in a membrane-in-the-middle (MIM) system \cite{thompson2008}. Quantum correlated light generated from a four-wave mixing (4WM) phase-insensitive amplifier in an atomic vapor cell \cite{McCormick:07,glorieux7727strong,Qin:12,Sheng:12} has attracted a lot of attention recently. Compared to the one generated by parametric down-conversion in a crystal \cite{PhysRevLett.59.2555}, it has several advantages, such as long coherence time, frequencies close to the atomic transitions, and multiple quantum correlated spatial modes. Such a unique class of quantum light source has been widely exploited in quantum imaging \cite{boyer2008entangled,marino2009tunable,Clark:12,PhysRevA.95.053849}, nonlinear interferometry \cite{hudelist2014quantum}, plasmonic sensing \cite{pooser2015plasmonic,Holtfrerich:16,dowran2018quantum}, and ultrasensitive measurement of microcantilever displacement in free space \cite{pooser2015ultrasensitive}. Very recently, the generation of squeezing below 10 Hz enables the possibility that the applications of twin beams can be extended to the low-frequency region \cite{wu2018twin}. 

\begin{figure}
\centering
\includegraphics[width=1\linewidth]{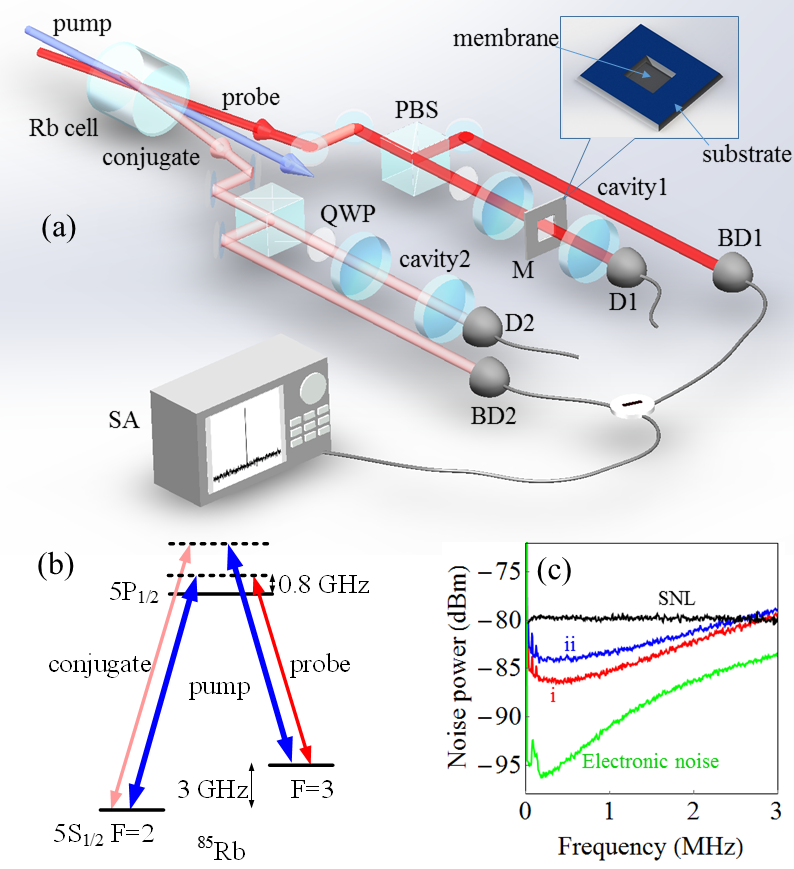}
\caption{(a) Schematic diagram of the experimental setup. PBS, polarization beam splitter; QWP, quarter-wave plate; M, membrane; BD1-BD2, Balanced photodetectors; D1-D2, photodetectors for cavity locking; SA, spectrum analyzer. The lenses for the purpose of mode-matching and focus are not displayed. The inset is the enlarged image of membrane with substrate. (b) The relevant energy level diagram of $^{85}$Rb D1 line showing a double-lambda configuration for the four-wave mixing process. (c) Noise power spectra. The black and green curves are the SNL and electronic noise floor, respectively. The red (i) and blue (ii) curves are the intensity difference squeezing between the probe and conjugate fields without and with the light beams reflecting from the cavities, respectively. The resolution bandwidth (RBW) and video bandwidth (VBW) are 10 $kHz$ and 1 $kHz$, respectively.}
\end{figure}

The displacement sensitivity on the order of ${\rm{fm}}/\sqrt {Hz}$ has been realized by combining multi-spatial-mode quantum light source with a differential measurement in free space \cite{pooser2015ultrasensitive}. However, the sensitivity with this method is difficult to be significantly further improved due to the limitation of the optical beam deflection measurement. Here, by placing the membrane in a medium-finesse optical cavity, the displacement measurement sensitivity of 200 $am/\sqrt {Hz}$ is achieved, which is greatly enhanced compared to a standard Michelson interferometer \cite{wu2018parametric} or a low-finesse optical cavity (see Supplementary Materials \cite{sup}). The measurement sensitivity is further improved with the application of quantum correlated light, and the noise floor can be squeezed 3 $dB$ below the SNL or even better, which depends on the cavity parameters. This quantum-correlated-light-assisted and cavity-based technique provides an alternative strategy for ultrasensitive displacement measurements with no need of a local oscillator and homodyne detection technique. This method is particularly useful for the measurements at weak light field, for example, the usable optical power is limited by the damage threshold of biological sample \cite{taylor2013biological}. This proof-of-principle displacement measurement of membrane should be applicable to other objects, such as atoms \cite{hood2000atom}, biological tissues \cite{taylor2013biological}, and strings \cite{anetsberger2009near}. Such an approach can also be straightforwardly extended to other systems for precision measurements, for instance, mass spectroscopy, accelerometers, and force sensors. In addition, a second optical cavity is added and the correlated frequency noise of narrow-band twin beams is directly analyzed. The measurement sensitivity is thereby optimized by minimizing the excess noise due to the classical laser frequency fluctuation.

\section{II. Experiment}

The quantum correlated beams are generated from a 4WM process by mixing the pump and probe beams in a hot rubidium (Rb) vapor cell \cite{McCormick:07,glorieux7727strong,Qin:12}, as shown in Fig. 1(a). Both the strong pump and weak probe beams are from the same diode laser. The probe beam is picked before going through the tapered amplifier and is tuned $\sim$ 3 $GHz$ to the red of the pump field by double passing through an acousto-optic modulator (AOM). The Rb vapor cell with anti-reflective coating has a length of 12 $mm$ and is held at $\sim$ 104 $^\circ$C without magnetic shielding. The pump and probe beams overlap inside the cell at a small angle ($\sim$ 0.3$^\circ$). The radii of the pump and probe beams are 550 $\mu m$ and 150 $\mu m$ at the center of the atomic cell, respectively. The pump beam is blue detuned $\sim$ 800 $MHz$ relative to the $^{85}Rb$ $5S_{1/2,} F=2 \to 5P_{1/2}, F'=3$ transition, as indicated in Fig. 1(b). The amplification gain of the probe beam is $\sim$ 7 with 300 $mW$ pump beam injecting into the cell.

The 4WM process is similar to the parametric down conversion phenomenon in nonlinear crystals, and the effective Hamiltonian of the generating twin beams can be written as $H = i\hbar \chi (a_0 a_p^\dag  a_c^\dag   - a_0^\dag  a_p a_c )$, where $a_i$ ($i=0,p,c$) describe the pump, probe and conjugate fields, respectively. This process is a third-order nonlinear effect, and the atoms convert two pump photons into one probe and one conjugate photon. The simultaneous generation of the probe and conjugate photon pairs ensures the quantum correlation between the amplitudes of twin beams, i.e. the fluctuations of the twin beams are identical even though the intensity of each beam may fluctuate. The quantum noise reduction is $1/(2g - 1)$, where $g$ is the probe gain of 4WM process. The intensity difference between the twin beams (the red curve) exhibits a noise reduction of $\sim$ 7 $dB$ below the SNL (the black curve) at the frequency of hundreds $kHz$ when the correlation of the probe and conjugate beams are measured directly out of the cell, as depicted in Fig. 1 (c).

The SNL is measured by splitting a coherent field with an equivalent total power of the probe and conjugate fields using a 50/50 beam splitter and sent to the balanced photodetector (BD1 and BD2). Since the coherent field obeys the Poisson distribution, the noise power after the subtracter is simply the sum of two channels of the balanced photodetector. The SNL can be determined by $V_{SNL}=\sqrt {2eRG^2 P\Delta f}$, where $e$ is the electron charge, $R=0.62$ $A/W$ is the responsivity of the photodiodes (corresponding to a quantum efficiency $\sim$ 95 $\%$ at 795 $nm$), $G=0.5 \times10^5$ $V/A$ is the transimpedance amplification gain of the photodetector, $P$ is the total power of the twin beams, and $\Delta f$ is the detection bandwidth. 

\begin{figure*}
\centering
\includegraphics[width=0.85\linewidth]{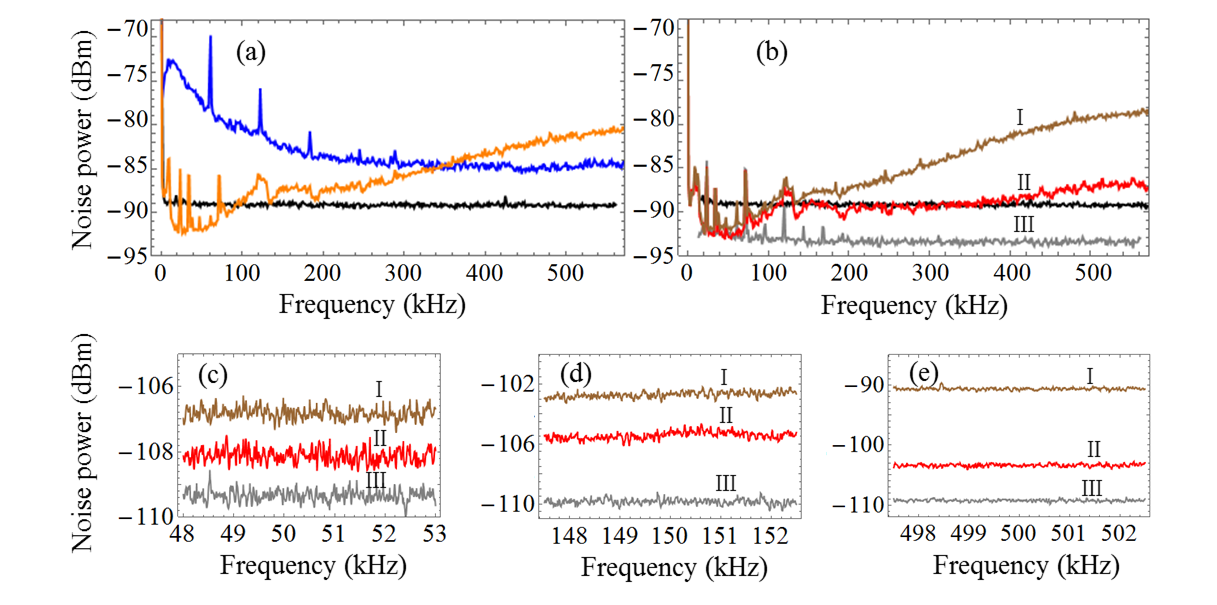}
\caption{Noise power spectra. (a) The blue and orange curves correspond to the noise power spectra when the laser linewidth is narrowed by locking to a stable cavity or not, respectively. In this case, the conjugate beam is directly shining on the BD2. The black curve is the SNL. (b) The brown (I) and red (II) curves represent that the frequency noise is added or subtracted, respectively. The gray (III) curve shows the case when the cavities are bypassed. 1 $kHz$ RBW and 100 $Hz$ VBW. (b-d) The narrow-spanned noise power spectra corresponding to Fig. 2(b) at frequencies of 50 $kH$z, 150 $kHz$, and 500 $kHz$, respectively. 30 $Hz$ RBW and 3 $Hz$ VBW. Each data trace is averaged over 50 measurements with identical parameters.}
\end{figure*}

To measure the vibration of the membrane, the probe beam is sent to the MIM cavity. The reflected probe beam from the cavity is collected on the BD1, and the power spectral density is measured by the electrical spectrum analyzer, as shown in Fig. 1(a). We use a 50 $nm$ stoichiometric SiN membrane. The reflection of the membrane at 795 $nm$ is $\sim$ 20$\%$, which is determined by the refractive index of SiN, the thickness of the membrane, and the wavelength of light field \cite{hecht1998optics}. The loss of membrane is extremely small ($<10^{-5}$). The finesse of the cavity is $\sim$ 2000. The cavity length, $l$, is $\sim$ 8.7 $cm$. The mode-matching efficiency of the cavity is intended to keep low, which is $\sim$ 10$\%$, therefore, most part of the light is reflected from the cavity in order to maintain the quantum correlations. During the measurement, the length of the cavity is stabilized via the side-peak lock technique. This is done by monitoring the cavity transmission with D1 and feedbacking the transmission signal to the PZT (not shown in Fig. 1(a)) where the cavity mirror is mounted on. The membrane's Brownian motion and the quantum back-action can be ignored compared to the SNL under the circumstance of atmospheric pressure and far off the mechanical resonances. Although the MIM system should be rigorously treated as two coupled cavities, we can adopt the modified two-mirror cavity model due to the large transmittance of membrane used in the experiment ($>80\%$) \cite{sup}.

The fluctuation of the membrane position leads to the amplitude modulation (AM) of the cavity reflection when the probe frequency is detuned from the cavity resonance. The sensitivity of the displacement measurement is determined by \cite{sup}
\begin{widetext}
\begin{equation}
\sqrt {S_{x,ref}^{shot} (\Omega )}  = \frac{\lambda }{{8\sqrt {2\eta _c P_{in} /(\hbar \omega )} \kappa ^2 |r_m |F}} \Bigg(\frac{{(\Omega ^2  + \Delta ^2  + \kappa ^2 )^2  - 4\Omega ^2 \Delta ^2 }}{{(\Omega ^2  + \kappa ^2 )/(\Delta ^2  + \kappa ^2 )}}\Bigg)^{1/2},
\end{equation}
\end{widetext}
where $\Omega$ is the analysis Fourier frequency, $F$ is the cavity finesse, $\lambda$ is the optical wavelength, ${r_m}$ is the membrane reflection coefficient, $\eta_c$ is quantum efficiency of the photodiode, $P_{in}$ is the cavity input power, $\Delta$ is the frequency detuning, and $\kappa$ is the cavity decay rate, respectively. In the slow modulation limit ($\Omega  <  < \kappa$), Eq. (1) can be interpreted by the cavity dispersion, i.e. the sensitivity is inversely proportional to the derivative of cavity reflection power to the frequency detuning. The steeper the slope of cavity reflection is, the larger AM will be caused by the small membrane position fluctuation, therefore, this leads to a better measurement sensitivity.

When the quantum correlated light is used instead of coherent light, the quantum noise is reduced by a factor $\beta = 1 - \eta  + \eta /(2g - 1)$ \cite{prl1990}, where the detection efficiency $\eta$ is included. Hence, the displacement measurement sensitivity with the quantum correlated light is given by
\begin{equation}
\sqrt {S_{x,ref}^{squeezed} (\Omega )}  = \sqrt {S_{x,ref}^{shot} (\Omega )\beta }.
\end{equation}

\section{III. Results and discussion}

To implement the precise measurement of membrane displacement, the noise floor is critically important. This requires paying particular attention to the quantum correlated light used in the experiment. The loss and other technical noises are serious issues in the sub-shot noise measurement. Noises under different situations are carefully characterized, as presented in Fig. 2. When the cavity is tuned near resonance with the probe beam, the noise background increases significantly, well above the SNL, as depicted by the blue curve in Fig. 2(a). This excess noise is owing to the laser frequency fluctuation, since both the cavity length and laser frequency modulation (FM) contribute equally to the AM of cavity reflection. We also performed the membrane-mirror cavity experiment and this excess noise is not observed, which is due to the much lower cavity finesse in this case compared to the MIM system \cite{sup}. 

This excess noise due to the FM-to-AM conversion hinders the quantum correlated light enhancement and degrades the measurement sensitivity. In the experiment, we adopt two ways to suppress the noise from the laser frequency fluctuation. Firstly, the laser linewidth is narrowed from 100 $kHz$ to 800 $Hz$ by locking the laser frequency to a high-finesse stable cavity, and the noise background substantially decreases, as shown by the orange curve in Fig. 2(a). This reduction is especially tremendous at low frequencies, which is more than 15 $dB$ reduction at 50 $kHz$. Secondly, we use the correlation of twin breams to further suppress the noise by introducing another cavity for the conjugate beam. The noise power of the sum and difference of twin beams exhibit different behaviors, as presented by the brown (I) and red (II) curves in Fig. 2(b).  Equally important, this provides a new method to exam the frequency correlation of twin beams. 

Intuitively, one might think that the frequency noise between the twin beam should be anti-correlated due to the correlation of their amplitudes. However, we experimentally find that the frequency noise are actually correlated. This conclusion is confirmed by carefully performing the same measurements with two identical coherent fields split by a beam splitter and comparing it with the case of twin beams. The correlation of frequency noises can be understood by considering that the source of probe frequency noise is mostly from the diode laser. The pump, probe, and conjugate fields need to satisfy the law of energy conservation, therefore, we expect that the probe frequency noise is also correlated with the conjugate frequency noise. Thus, the cavity transfer functions for the two beams are necessary to have the same sign in order to cancel the frequency noise, which can be realized by locking the probe and conjugate cavities (cavity 1 and 2 in Fig. 1(a), respectively) on the same slope sides, where we obtain the red curve (II) in Fig. 2(a). When the probe and conjugate cavities are locked on the opposite sides of the cavity resonance, the frequency noise of the twin beams is actually added, which is exhibited as the brown curve (I). 

Although we carefully balance the parameters of the two cavities, the frequency noise cannot be completely suppressed, otherwise, the noise floor should approach to the gray curve (III) in Fig. 2(b), which is the case when both cavities are bypassed, i.e. the frequency noise has no contribution. In this case, the quantum noise reduction becomes $\sim$ 4.5 $dB$ instead of initial 7 $dB$, due to the excess optical loss from the PBS, mirrors, and window (the cavities are sealed inside a chamber for the purpose of vibration isolation). This indicates that the total detection efficiency reduces from to $87\%$ to $70\%$ due to the excess loss. The noise cannot be completely canceled at relatively large frequency because the laser frequency noise is relatively large at those frequencies, the two cavities are not exactly the same, and noise response might not be perfectly the same.

In order to better show the effect of noise suppression, Figs. 2(c-e) give the narrow-spanned noise power spectra of Fig. 2(b) at frequencies of 50 $kHz$, 150 $kHz$, and 500 $kHz$, respectively. Here, the SNL is $\sim$ 105 $dBm$ with 30 $H$z RBW. Generally, we find that the smaller the frequency noise is, the better noise reduction can be achieved. When the frequency noise is large, then the difference between the summation and subtraction of frequency noise of the twin beams is more significant.

\begin{figure}
\centering
\includegraphics[width=1\linewidth]{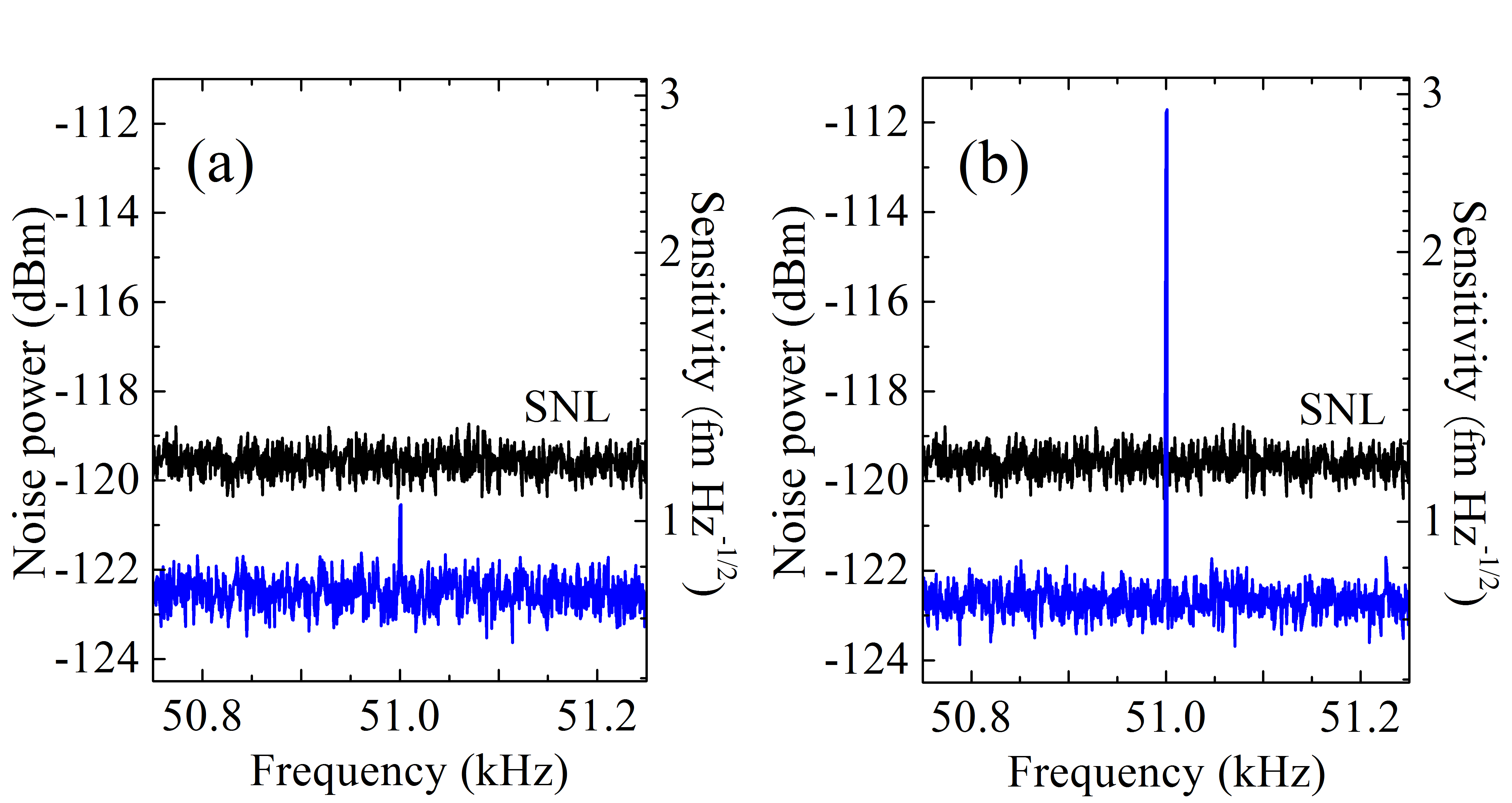}
\caption{Measured power spectra with twin beams. (a) $\Delta L=1$ $fm/\sqrt {Hz}$ and (b) $\Delta L=3$ $fm/\sqrt {Hz}$, respectively. 1 $Hz$ RBW and 1 $Hz$ VBW. Each data trace is averaged over 200 measurements with identical parameters. 
}
\label{fig3}
\end{figure}

We now demonstrate the twin-beam-enhanced sensitivity of the membrane displacement measurement. The vibration of membrane is excited by electrically driving the piezoelectric transducer (PZT) where the membrane is glued on, and hence, the displacement of the membrane can be controlled by tuning the voltage applied on the PZT. A Michelson interferometer is utilized to calibrate its response. With this calibrated PZT, the relationship between the displacement and the signal strength measured in the MIM system can be determined. The measurement sensitivity with twin beams is shown in Fig. 3, where the sensitivity is defined as the displacement power spectral density when the signal-to-noise ratio (SNR) is one. Please note that the noise is the lowest reachable noise floor rather than the SNL. The noise floor is reduced from $1.16 \pm 0.03$ $fm/\sqrt {Hz}$ (black curve) to $0.83 \pm 0.03$ $fm/\sqrt {Hz}$ (blue curve), which corresponds to a 95 $\%$ increase in the SNR or 40 $\%$ of the displacement measurement sensitivity. The recorded difference spectra give a 3 $dB$ noise reduction below the SNL around 51 $kHz$ with the optimized noise suppression. It is obvious that when the displacement of membrane is small, the signal will be buried in the SNL with coherent light, while a good SNR can still be achieved with twin beams, as shown in Fig. 3(a). 

\begin{figure}
\centering
\includegraphics[width=0.85\linewidth]{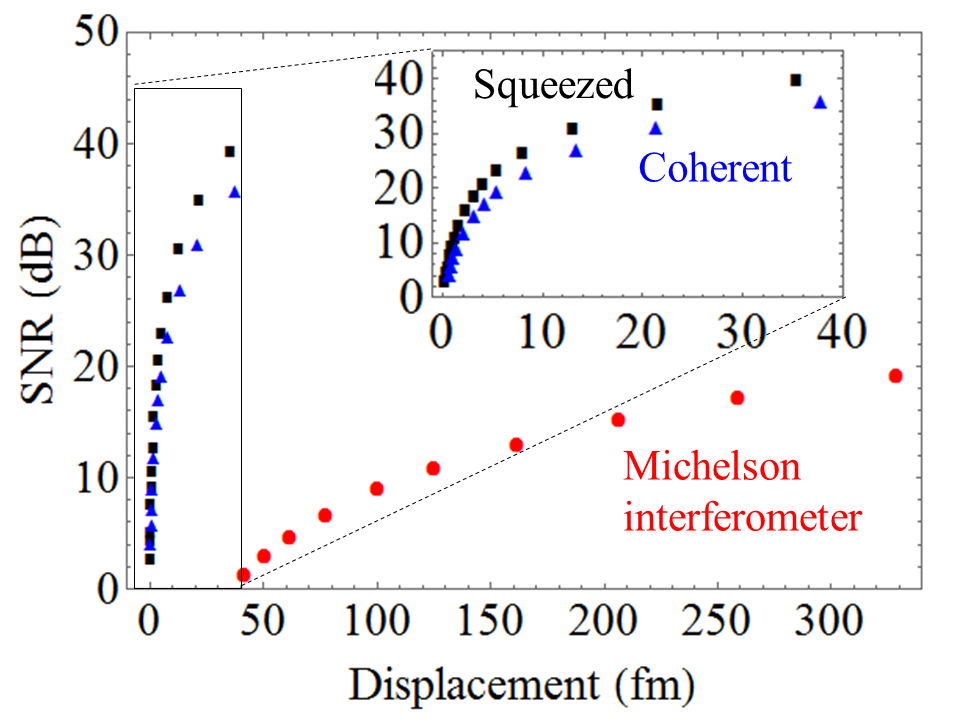}
\caption{SNR as a function of displacement amplitude. The black squares and blue triangles correspond to the measurements with quantum correlated light and coherent light, respectively. The red dots represent the Michelson interferometer measurement with the coherent light of equivalent power. The inset is the zoomed graph that displays the displacement range of $0-40$ $fm$. The membrane vibrates at frequency of 51 $kHz$. 1 $Hz$ RBW and 1 $Hz$ VBW.
}
\label{fig4}
\end{figure}

Figure 4 illustrates the SNR as a function of the displacement amplitude of the membrane at 51 $kHz$. For comparison, besides the measurements with quantum correlated light (black squares), the displacement measurement of membrane with coherent light and Michelson interferometer are also implemented in the experiment, as shown by the blue triangles and red dots in Fig. 4, respectively. For these three sets of experiments, the total input optical power is fixed to be 100 $\mu W$. The MIM scheme investigated here dramatically enhances the SNR compared to the Michelson interferometer measurement, and the sensitivity is two orders of magnitude better. The measurement with the quantum correlated light has higher SNR than with the coherent light at the equivalent optical power. 

At last, we investigate the SNR as a function of the cavity round trip phase $\phi$. The cavity round trip phase is controlled by locking the cavity at different frequencies, i.e., changing the frequency detuning. At different phases, the slope of cavity reflection, i.e. the length-to-power transfer function, is correspondingly changed and this leads to different signal enhancements. The dependence of SNR on the phase is shown in Fig. 5. The blue triangles and red dots in Fig. 5(a) correspond to the cases when the conjugate cavity is bypassed or not, respectively. The displacement sensitivity is improved by using the conjugate cavity. The inset indicates the noise power spectra of the data points at $\phi$ = -1.2 $mrad$. The blue and red curves are corresponding to the blue triangle and red dot, and the black curve is the SNL. Although the noise floor is squeezed by 3 $dB$ with the conjugate cavity, by comparing the blue and red curves, the noise is still larger than the SNL. The noise floor can be reduced below the SNL only when the locking position is not close to the maximum slope. To fully optimize the cavity performance, it requires a laser source with lower frequency noise, e.g. a Ti-sapphire laser. The best sensitivity $200\,am/\sqrt{Hz}$ is achieved in the current experimental parameters. 
Figure 5(b) is the theoretical simulations based on the modified two-mirror cavity model with the experimental parameters, where the finesse of the bare cavity and the input laser power are chosen to be 2000 and 100 $\mu W$. The solid curve is the sensitivity and the dashed curve is the cavity reflection. The observed improvement is consistent with the theoretical calculation by considering the measured losses and mode-matching effect.

\begin{figure}
\centering
\includegraphics[width=1\linewidth]{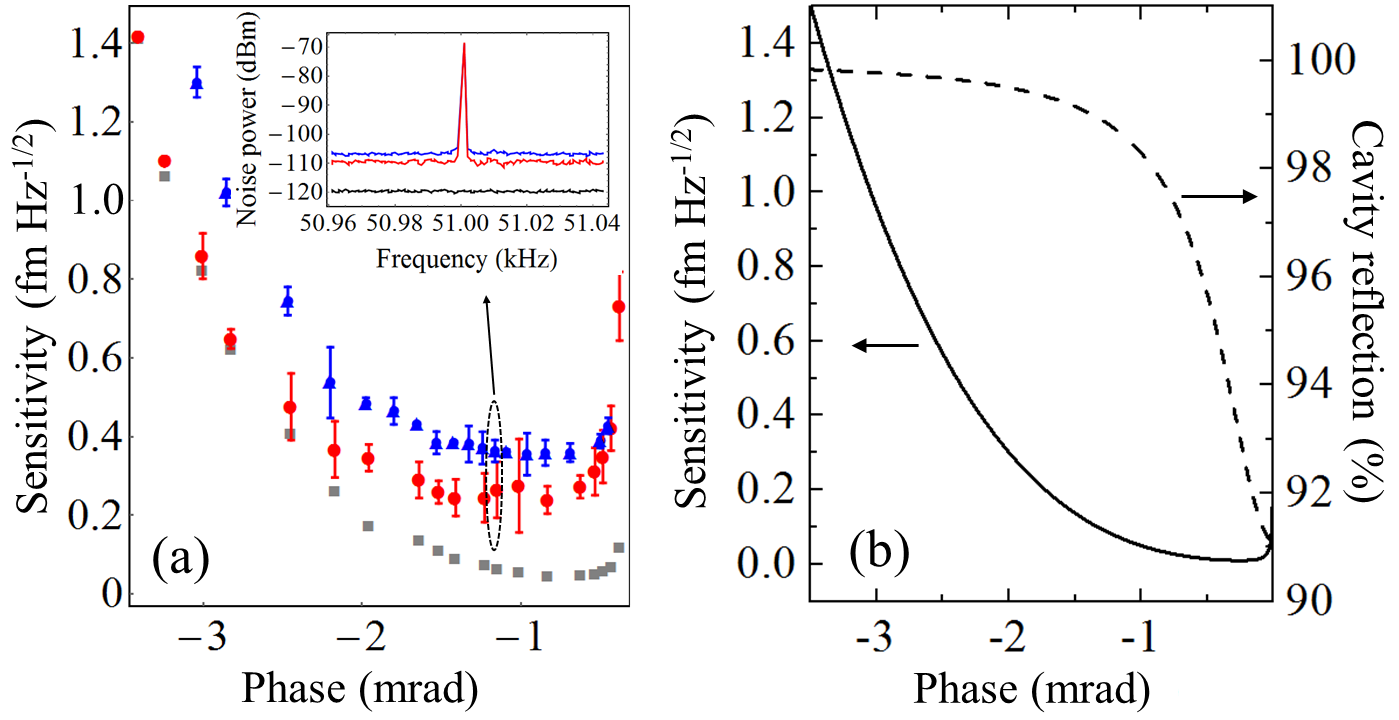}
\caption{Dependence of displacement sensitivity on the cavity round trip phase. (a) The measurement sensitivity as a function of the cavity locking position at 51 $kHz$. The blue triangles present the case when the conjugate cavity is bypassed, i.e. the conjugate beam is directly shining on BD2. The red dots are when the conjugate cavity is utilized to reduce the laser frequency noise. The gray squares stand for the ideal case when the frequency noise can be completely canceled, i.e. the noise level is 4.5 $dB$ below the SNL. The inset displays the noise power spectra of the corresponding data points. The blue and red curves display the blue triangle and red dot at $\phi$ = -1.2 $mrad$, respectively. 1 $Hz$ RBW and 1 $Hz$ VBW. (b) Theoretical calculation. The solid curve is the sensitivity and the dashed curve is the cavity reflection.
}
\end{figure}

\section{IV. Conclusion}

In conclusion, the sensitivity of displacement measurement has been greatly enhanced with a precision below the SNL by combining the optical cavity and quantum correlated light. Equally important, the correlation of the twin beams are directly investigated by using two independent cavities. The technique noises are substantially suppressed by subtracting the reflection of the two cavity. This method therefore reduces the constrain of laser light sources, since classical amplitude noise can be eliminated by the differential detection and frequency noise can be suppressed by using two cavities. This proof-of-principle demonstration should be able to extend to other systems where quantum back-action dominates the uncertainty. It also offers a new scheme to study cavity optomechanics, for instance, investigating correlation and entanglement of multiple membranes with quantum correlated light.

\section*{ACKNOWLEDGMENTS}
JS gratefully acknowledges Ashok Kumar for helpful discussions. This research is supported by the National Key Research and Development Program of China (2017YFA0304201), National Natural Science Foundation of China (11374101, 91536112, 11704126), Natural Science Foundation of Shanghai (17ZR1443100), the Shanghai Sailing Program (17YF1403900), the Program for Professor of Special Appointment (Eastern Scholar) at Shanghai Institutions of Higher Learning.

\vskip 0.1in

$^*$jtsheng@lps.ecnu.edu.cn

$^\dag$hbwu@phy.ecun.edu.cn

\bibliographystyle{apsrev4-1}

\end{document}